\documentstyle[twoside,fleqn,espcrc2,epsfig]{article}

\newcommand{\Z}{{\sf Z \!\!\! Z}}

\newcommand{\Sign}{\mbox{Sign}}
\newcommand{\PP}{\overline{\Psi} \Psi}
\newcommand{\up}{\uparrow}
\newcommand{\down}{\downarrow}

\title{Meron-Cluster Solution of Fermion and Other Sign Problems}

\author{J. Cox, C. Gattringer, K. Holland, B. Scarlet and U.-J. Wiese 
\address{Center for Theoretical Physics, Laboratory for Nuclear Science and 
Department of Physics \\ 
Massachusetts Institute of Technology (MIT), Cambridge, MA 02139}}

\begin{document}

\begin{abstract}

Numerical simulations of numerous quantum systems suffer from the notorious
sign problem. Important examples include QCD and other field theories at
non-zero chemical potential, at non-zero vacuum angle, or with an odd number of
flavors, as well as the Hubbard model for high-temperature superconductivity 
and quantum antiferromagnets in an external magnetic field. In all these cases
standard simulation algorithms require an exponentially large statistics in
large space-time volumes and are thus impossible to use in practice. 
Meron-cluster algorithms realize a general strategy to solve severe sign 
problems but must be constructed for each individual case. They lead to a 
complete solution of the sign problem in several of the above cases.

\end{abstract}

\maketitle

\section{Introduction}

Sign problems prevent the numerical solution of several important problems in 
physics. For example, QCD and other field theories at non-zero chemical 
potential or at non-zero vacuum angle have a complex action and hence complex
Boltzmann weights which cannot be interpreted as probabilities in a Monte Carlo
calculation. When the complex phase of the Boltzmann factor is included in
measured observables, the fluctuations of the phase give rise to dramatic 
cancellations. Especially for large systems at low temperatures this leads to 
relative statistical errors that are exponentially large in both the volume and
the inverse temperature. Similarly, the minus-signs that arise as a consequence
of Fermi statistics cause severe sign problems in numerical simulations of 
strongly correlated electron systems such as the Hubbard model for
high-temperature superconductors. Another class of systems that suffer from a 
sign problem are frustrated quantum spin systems --- for example, quantum
antiferromagnets in an external magnetic field. In all these cases it is
impossible in practice to study these systems with standard numerical methods.

Recently, it has been shown to be possible to completely solve some severe sign
problems using 
meron-cluster algorithms \cite{Cha99a,Cha99b}. The solution of the problem 
proceeds in two steps. The idea of the first step is to decompose a field 
configuration into independent clusters whose flip affects the sign in a 
definite way. Clusters whose flip changes the sign are referred to as merons.
The flip of a meron-cluster leads to an exact cancellation between two 
contributions $\pm 1$. Using an improved estimator, this reduces the problem of
canceling many contributions $\pm 1$ to the problem of averaging over 
non-negative contributions $0$ and $1$. An algorithm based on this first step 
has been successfully applied to the simulation of a bosonic model with a 
complex action --- the 2-d $O(3)$ model at non-vanishing $\theta$-vacuum angle 
\cite{Bie95}. In this model the sign-changing clusters are half-instantons ---
thus the name meron. However, the improved estimator alone solves only one half
of the sign problem because most of the time one generates contributions of 0
to the average sign and very rarely one encounters a contribution of $1$. In 
order to solve the other half of the problem a second step is necessary
which guarantees that contributions of $0$ and $1$ are generated with similar 
probabilities. The idea behind the second step is to include a Metropolis 
decision in the process of cluster decomposition. The two basic ideas behind 
the algorithm are general and can be applied to a variety of systems. However,
even though the meron concept is universal, the meron-cluster algorithm must be
constructed for each individual case. Here we present several cases, from both 
particle and condensed matter physics, in which a meron-cluster algorithm
has led to a complete solution of a severe sign problem.

\section{General Nature of the Sign Problem}

First of all, it should be stressed that the sign problem is strongly 
influenced by the choice of basis for the physical Hilbert space. When one 
constructs a path integral representation of a quantum statistical partition 
function 
\begin{equation}
Z = \mbox{Tr} \exp(- \beta H), 
\end{equation}
one divides the Euclidean time interval of extent $\beta$ into $M$ small 
time-steps of size $\epsilon$ (with $\beta = M \epsilon$) and one inserts 
complete sets of basis states $|n\rangle$ between the operators 
$\exp(- \epsilon H)$. The product of the resulting transfer matrix elements 
$\langle n|\exp(- \epsilon H)|n'\rangle$ defines the Boltzmann factor. 
Depending on the choice of basis $|n\rangle$, the Boltzmann weight may be
positive, negative or even complex. When the Boltzmann weight is complex, one
can always restrict oneself to the real part, because the total partition
function is always real (as long as $H$ is Hermitean). Still, the Boltzmann
weight may be negative and in general it takes the form $\Sign[n] 
\exp(- S[n])$, with $\Sign[n] = \pm 1$ and $\exp(- S[n]) \geq 0$. In principle,
it is always possible to avoid the sign problem (i.e. one can always ensure 
that $\Sign[n] = 1$) by a clever choice of basis $|n\rangle$ of the physical 
Hilbert space. For example, when one chooses the basis of Hamiltonian 
eigenstates, $H |n\rangle = E_n |n\rangle$, all transfer matrix elements 
$\langle n|\exp(- \epsilon H)|n'\rangle = \exp(- \epsilon E_n) \delta_{n,n'}$ 
and hence the Boltzmann weights are non-negative. Of course, this is a rather
academic solution of the sign problem, because the cases we want to simulate 
are the ones for which we don't know how to diagonalize the Hamiltonian. Still,
the argument shows that a change of basis in the path integral may have a large
impact on the sign problem.

Let us now consider a general path integral 
\begin{equation}
Z = \sum_n \Sign[n] \exp(-S[n])
\end{equation}
over configurations $n$ with a Boltzmann weight of $\Sign[n] = \pm 1$ and 
magnitude $\exp(- S[n])$. Here $S[n]$ is the action of a modified model, with 
partition function $Z' = \sum_n \exp(- S[n])$, which does not suffer from the
sign problem and which can thus be simulated with standard Monte Carlo methods.
An observable $O[n]$ of the original model is obtained in a simulation of the 
modified ensemble as
\begin{eqnarray}
\langle O \rangle = \frac{1}{Z} \sum_n O[n] \Sign[n] \exp(- S[n]) =
\frac{\langle O \ \Sign \rangle}{\langle \Sign \rangle}. \nonumber \\ \,
\end{eqnarray}
The average sign in the modified ensemble is given by
\begin{eqnarray}
\langle \Sign \rangle&=&\frac{1}{Z'}{\sum_n \Sign[n] \exp(-S[n])} \nonumber \\
&=&\frac{Z}{Z'} = \exp(- \beta V \Delta f).
\end{eqnarray}
The last equality points to the heart of the sign problem. The expectation 
value of the sign is exponentially small in both the volume $V$ and the inverse
temperature $\beta$ because the difference between the free energy densities 
$\Delta f = f - f'$ of the original and the modified systems is necessarily 
positive.

Even in an ideal simulation of the modified ensemble which generates $N$ 
completely uncorrelated configurations, the relative statistical error of the 
sign is
\begin{equation}
\frac{\Delta \Sign}{\langle \Sign \rangle} = 
\frac{\sqrt{\langle \Sign^2 \rangle - \langle \Sign \rangle^2}}
{\sqrt{N} {\langle \Sign \rangle}} = \frac{\exp(\beta V \Delta f)}{\sqrt{N}}.
\end{equation}
Here we have used that $\Sign^2 = 1$. To determine the average sign with 
sufficient 
accuracy one needs to generate on the order of $N = \exp(2 \beta V \Delta f)$ 
configurations. For large volumes and small temperatures this is impossible in
practice. It is possible to solve one half of the problem if one can match any 
contribution $-1$ with another contribution $1$ to give $0$, such that only a 
few unmatched contributions $1$ remain. Then effectively $\Sign = 0,1$ and 
hence $\Sign^2 = \Sign$. This reduces the relative error to
\begin{equation}
\frac{\Delta \Sign}{\langle \Sign \rangle} = 
\frac{\sqrt{\langle \Sign \rangle - \langle \Sign \rangle^2}}
{\sqrt{N'} {\langle \Sign \rangle}} = 
\frac{\exp(\frac{\beta V \Delta f}{2})}{\sqrt{N'}}.
\end{equation}
One gains an exponential factor in statistics, but one still needs to generate 
$N' = \sqrt{N} = \exp(\beta V \Delta f)$ independent configurations in order to
determine the average sign accurately. This difficulty still arises because one
generates 
exponentially many vanishing contributions before one encounters a contribution
$1$. As explained below, in the meron-cluster algorithm an explicit matching of
contributions $-1$ and $1$ is achieved using an improved estimator. This step
solves
one half of the sign problem. In a second step involving a Metropolis decision,
the algorithm ensures that contributions $0$ and $1$ occur with similar 
probabilities. This step saves another exponential factor in statistics and 
solves the other half of the sign problem.

\section{Chiral Phase Transition with Staggered Fermions $\,^2$}
\footnotetext[2]{Based on a talk presented by K. Holland}

To illustrate the power of the meron-cluster algorithm, we apply it to a model
of $(3+1)$-d staggered fermions in the Hamiltonian formulation. The model has
$N = 2$ flavors and a $\Z(2)$ chiral symmetry that is spontaneously broken at 
low temperatures. The fermion determinant can be negative in this model. Hence,
due to the sign problem, standard fermion algorithms fail in this case. The
meron-cluster algorithm is the only numerical method available to simulate this
model. In our method we do not integrate out the fermions but describe them in 
a Fock state basis. The resulting bosonic model of fermion occupation numbers 
interacts locally, but has a non-local fermion permutation sign resulting from 
the Pauli exclusion principle. Standard numerical methods would suffer from 
severe cancellations of positive and negative contributions to the partition 
function. Like other cluster methods, the meron-cluster algorithm decomposes a 
configuration of fermion occupation numbers into clusters which can be flipped 
independently. Under a cluster flip an occupied site becomes empty and vice 
versa. The main idea of the meron-cluster algorithm is to construct the 
clusters such that they affect the fermion sign independently when they are 
flipped. In addition, it must always be possible to flip the clusters into a 
reference configuration with a positive sign. Like other cluster algorithms, 
the meron algorithm substantially reduces critical slowing down. This advantage
allows us to work directly in the chiral limit. 

We consider staggered fermions hopping on a 3-d cubic spatial lattice with 
$V = L^3$ sites $x$ ($L$ even) and with periodic or antiperiodic spatial 
boundary conditions. We start in the Hamiltonian formulation and then derive a 
path integral on a $(3+1)$-d Euclidean space-time lattice. The fermions are 
described by creation and annihilation operators $\Psi_x^+$ and $\Psi_x$ with 
standard anticommutation relations
\begin{equation}
\{\Psi_x^+,\Psi_y^+\} = \{\Psi_x,\Psi_y\} = 0, \, 
\{\Psi_x^+,\Psi_y\} = \delta_{xy}.
\end{equation}
The staggered fermion Hamilton operator takes the form 
\begin{equation}
H = \sum_{x,i} h_{x,i} + m \sum_x (-1)^{x_1+x_2+x_3} \Psi_x^+ \Psi_x
\end{equation}
that is a sum of nearest-neighbor couplings $h_{x,i}$ and a mass term $m \PP$. 
From here on, we work directly in the chiral limit $m = 0$, and only use 
$\PP$ as an observable. The term $h_{x,i}$ couples the fermion operators at the
lattice sites $x$ and $x+\hat i$, where $\hat i$ is a unit-vector in the 
$i$-direction, and
\begin{eqnarray}
h_{x,i}&=&\frac{1}{2} \eta_{x,i}(\Psi_x^+ \Psi_{x+\hat i} + 
\Psi_{x+\hat i}^+ \Psi_x) \nonumber \\
&+&G (\Psi_x^+ \Psi_x - \frac{1}{2})
(\Psi_{x+\hat i}^+ \Psi_{x+\hat i} - \frac{1}{2}).
\end{eqnarray}
Here $\eta_{x,1} = 1$, $\eta_{x,2} = (-1)^{x_1}$ and $\eta_{x,3} = 
(-1)^{x_1 + x_2}$ are the standard staggered fermion sign factors and $G$ is a
four-fermion coupling constant. 

To construct a path integral for the partition function, we decompose the 
Hamilton operator into six terms $H = H_1 + H_2 + ... + H_6$ with
\begin{equation}
H_i = \!\!\! \sum_{\stackrel{x = (x_1,x_2,x_3)}{x_i even}} \!\!\! h_{x,i}, 
\, \,
H_{i+3} = \!\!\! \sum_{\stackrel{x = (x_1,x_2,x_3)}{x_i odd}} \!\!\! h_{x,i}.
\end{equation}
The individual contributions to a given $H_i$ commute with each other, but two 
different $H_i$ do not commute. Using the Suzuki-Trotter formula, we express 
the fermionic partition function at inverse temperature $\beta$ as
\begin{eqnarray}
Z&=&\mbox{Tr} \exp(- \beta H) \nonumber \\
&=&\lim_{M \rightarrow \infty} \mbox{Tr} 
[\exp(- \epsilon H_1) ... \exp(- \epsilon H_6)]^M.
\end{eqnarray}
We have introduced $6M$ Euclidean time slices with $\epsilon = \beta/M$ being 
the lattice spacing in the Euclidean time direction. Following Jordan and 
Wigner \cite{Jor28} we represent the fermion operators by Pauli matrices
\begin{eqnarray}
\Psi_x^+&=&\sigma_1^3 \sigma_2^3 ... \sigma_{l-1}^3 \sigma_l^+, \,
\Psi_x = \sigma_1^3 \sigma_2^3 ... \sigma_{l-1}^3 \sigma_l^-, \nonumber \\
n_x&=&\Psi_x^+ \Psi_x = \frac{1}{2}(\sigma_l^3 + 1),
\end{eqnarray}
with
\begin{equation}
\sigma_l^\pm = \frac{1}{2} (\sigma_l^1 \pm i \sigma_l^2), \, 
[\sigma_l^i,\sigma_m^j] = 2 i \delta_{lm} \epsilon_{ijk} \sigma_l^k.
\end{equation}
Here $l$ labels the lattice point $x$. 

We now insert complete sets of fermion Fock states between the factors 
$\exp(- \epsilon H_i)$. Each site is either empty or occupied, i.e. $n_x$ has 
eigenvalue $0$ or $1$. In the Pauli matrix representation this corresponds to 
eigenstates $|0\rangle$ and $|1\rangle$ of $\sigma_l^3$ with 
$\sigma_l^3 |0\rangle = - |0\rangle$ and $\sigma_l^3 |1\rangle = |1\rangle$. 
The transfer matrix is a product of factors
\begin{eqnarray}
\label{transfer}
&&\exp(- \epsilon h_{x,i}) = \exp(\frac{\epsilon G}{4}) \nonumber \\
&&\times \left(\begin{array}{cccc}
\exp(- \frac{\epsilon G}{2}) & 0 & 0 & 0 \\ 
0 & \cosh \frac{\epsilon}{2} & \Sigma \sinh \frac{\epsilon}{2} & 0 \\ 
0 & \Sigma \sinh \frac{\epsilon}{2} & \cosh \frac{\epsilon}{2} & 0 \\ 
0 & 0 & 0 & \exp(- \frac{\epsilon G}{2}) \end{array} \right),
\nonumber \\ \
\end{eqnarray}
which is a $4 \times 4$ matrix in the Fock space basis $|00\rangle$,
$|01\rangle$, $|10\rangle$ and $|11\rangle$ of two sites $x$ and $x+\hat i$.
Here $\Sigma = \eta_{x,i} \sigma_{l+1}^3 \sigma_{l+2}^3 ... \sigma_{m-1}^3$ 
includes the local sign $\eta_{x,i}$ as well as a non-local string of Pauli 
matrices running over consecutive labels between $l$ and $m$, where $l$ labels 
the lattice point $x$ and $m$ labels $x+\hat i$. Note that $\Sigma$ is diagonal
in the occupation number basis.

The partition function is now expressed as a path integral 
$Z = \sum_n \Sign[n] \exp(- S[n])$ over configurations of occupation numbers 
$n(x,t) = 0,1$ on a $(3+1)$-d space-time lattice of points $(x,t)$. The 
Boltzmann factor is a product of space-time plaquette contributions with an
action $S[n(x,t),n(x+\hat i,t),n(x,t+1),n(x+\hat i,t+1)]$ and with
\begin{eqnarray}
\label{Boltzmann}
&&e^{- S[0,0,0,0]} = e^{- S[1,1,1,1]} = e^{- \epsilon G/2},
\nonumber \\
&&e^{- S[0,1,0,1]} = e^{- S[1,0,1,0]} = \cosh \frac{\epsilon}{2},
\nonumber \\
&&e^{- S[0,1,1,0]} = e^{- S[1,0,0,1]} = \sinh \frac{\epsilon}{2}.
\end{eqnarray}
All the other Boltzmann weights are zero.

The sign of a configuration, $\Sign[n]$, also is a product of space-time 
plaquette contributions
$\Sign[n(x,t),n(x+\hat i,t),n(x,t+1),n(x+\hat i,t+1)]$ with
\begin{eqnarray}
&&\Sign[0,0,0,0]) = \Sign[0,1,0,1]) = \nonumber \\
&&\Sign[1,0,1,0]) = \Sign[1,1,1,1]) = 1, \nonumber \\
&&\Sign[0,1,1,0]) = \Sign[1,0,0,1]) = \Sigma.
\end{eqnarray}
It should be noted that $\Sigma$ gets contributions from all lattice points 
with labels between $l$ and $m$. This seems to make an evaluation of the 
fermion sign rather tedious. Also, it is not a priori obvious that $\Sign[n]$ 
is independent of the arbitrarily chosen order of the lattice points. 
Fortunately, there is a simple way to compute $\Sign[n]$, which is directly 
related to the Pauli exclusion principle and is manifestly 
order-independent. In fact, $\Sign[n]$ has a topological meaning. The occupied 
lattice sites define fermion world-lines which are closed around the Euclidean 
time direction. Of course, during their Euclidean time evolution fermions can 
interchange their positions and the fermion world-lines define a permutation 
of particles. The Pauli exclusion principle dictates that the fermion sign is 
just the sign of that permutation. If we work with antiperiodic spatial 
boundary conditions, $\Sign[n]$ receives an extra minus-sign for every fermion 
world-line that crosses a spatial boundary. Figure 1 shows two configurations 
of fermion occupation numbers in $(1+1)$ dimensions. The first configuration 
corresponds to two fermions at rest and has $\Sign[n] = 1$. In the second 
configuration two fermions interchange their positions with one fermion 
stepping over the spatial boundary. If one uses periodic spatial boundary 
conditions this configuration has $\Sign[n] = - 1$. Note that the same 
configuration would have $\Sign[n] = 1$ when antiperiodic boundary conditions 
are used.
\begin{figure}[htb]
\epsfig{file=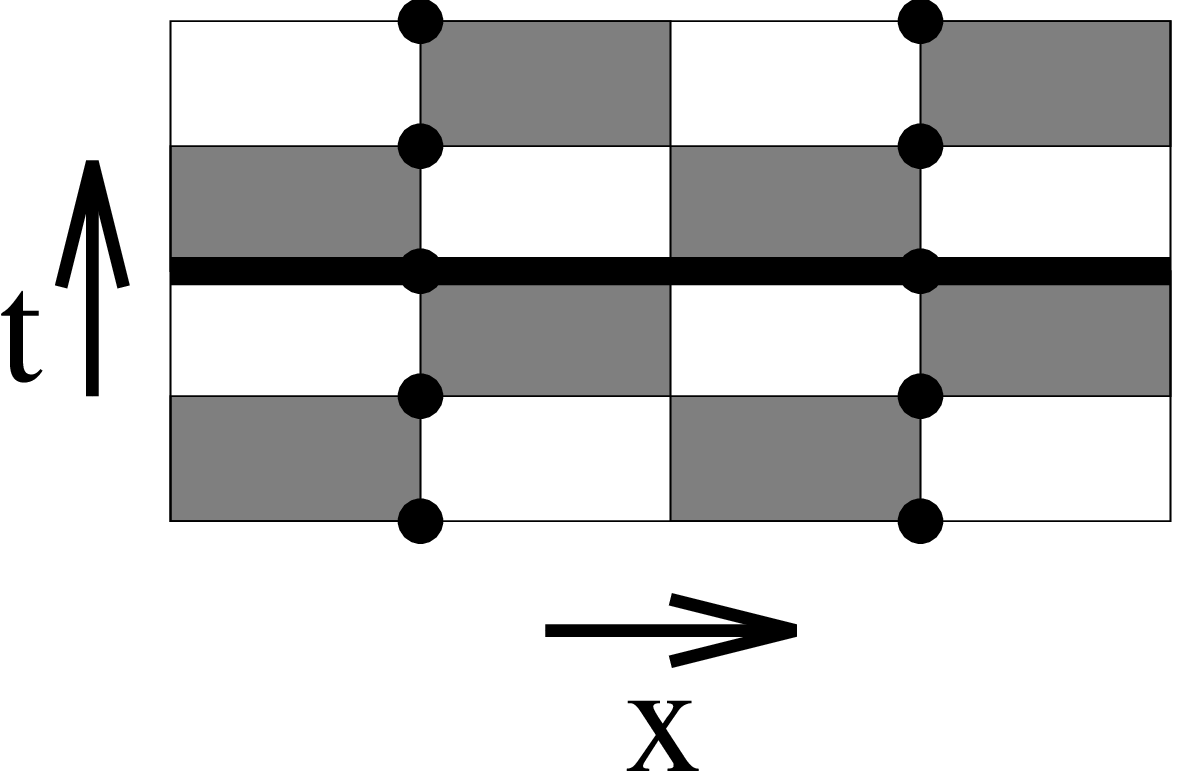,
height=5.5cm,angle=270,
bbllx=0,bblly=0,bburx=221,bbury=287}
\epsfig{file=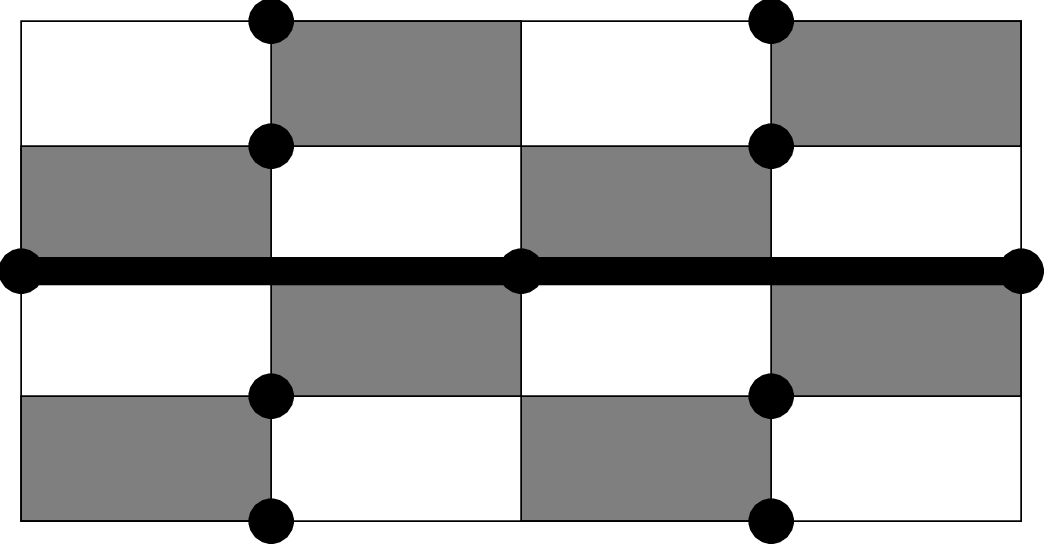,
height=5.5cm,angle=270,
bbllx=0,bblly=-43,bburx=157,bbury=250}
\vspace{-0.5cm}
\caption{\it Two configurations of fermion occupation numbers in $(1+1)$ 
dimensions. The shaded plaquettes carry the interaction. The dots represent 
occupied sites. In the second configuration, two fermions interchange their 
positions. The fat line represents a meron-cluster. Flipping this cluster 
changes one configuration into the other and changes the fermion sign.}
\end{figure}

Quantities of physical interest are the chiral condensate
\begin{equation}
\PP[n] = \frac{\epsilon}{6} \sum_{x,t} (-1)^{x_1 + x_2 + x_3} 
(n(x,t) - \frac{1}{2})
\end{equation}
and the corresponding chiral susceptibility
\begin{equation}
\chi = \frac{1}{\beta V} \langle (\overline{\Psi} \Psi)^2 \rangle_f.
\end{equation}

Up to now we have derived a path integral representation for the fermion
system in terms of bosonic occupation numbers and a fermion sign factor that
encodes Fermi statistics. The system without the sign factor is bosonic and
is characterized by the positive Boltzmann factor $\exp(- S[n])$. Here the 
bosonic model is a quantum spin system with the Hamiltonian
\begin{equation}
H = \sum_{x,i} (S_x^1 S_{x+\hat i}^1 + S_x^2 S_{x+\hat i}^2 + 
G S_x^3 S_{x+\hat i}^3), 
\end{equation}
where $S_x^i = \frac{1}{2} \sigma_l^i$ is a spin $1/2$ operator associated with
the lattice site $x$ that was labeled by $l$. From here on, we restrict
ourselves to $G = 1$, which corresponds to the antiferromagnetic quantum 
Heisenberg model. In the language of the spin model, the chiral condensate 
turns into the staggered magnetization
\begin{equation}
\PP = \frac{\epsilon}{6} \sum_{x,t} (-1)^{x_1+x_2+x_3} S_x^3.
\end{equation}
The meron-cluster fermion algorithm is based on a cluster algorithm for the 
corresponding bosonic model without the sign factor. Bosonic quantum spin 
systems can be simulated very efficiently with cluster algorithms 
\cite{Wie92,Eve93,Eve97}. The first cluster algorithm for lattice fermions was 
described in \cite{Wie93}. These algorithms can be implemented directly in the 
Euclidean time continuum \cite{Bea96}, i.e. the Suzuki-Trotter discretization 
is not even necessary. The decomposition of the lattice into clusters results 
from connecting neighboring sites on each individual space-time interaction 
plaquette following probabilistic cluster rules. A sequence of connected sites 
defines a cluster. In this case the clusters are sets of closed loops. The 
cluster rules are constructed to obey detailed balance. To show this constraint
we write the plaquette Boltzmann factors as
\begin{eqnarray}
\label{cluster}
&&\!\!\!e^{- S[n(x,t),n(x+\hat i,t),n(x,t+1),n(x+\hat i,t+1)]} = \nonumber \\
&&\!\!\!A \delta_{n(x,t),n(x,t+1)} \delta_{n(x+\hat i,t),n(x+\hat i,t+1)} +
\nonumber \\
&&\!\!\!B \delta_{n(x,t),1-n(x+\hat i,t)} \delta_{n(x,t+1),1-n(x+\hat i,t+1)}.
\end{eqnarray}
The $\delta$-functions specify which sites are connected and thus belong to the
same cluster. The quantities $A$ and $B$ determine the relative probabilities 
for different cluster break-ups of an interaction plaquette. Inserting the 
expressions from eq.(\ref{Boltzmann}) one finds
\begin{equation}
\label{balance}
A = e^{- \epsilon/2}, B = \sinh \frac{\epsilon}{2}.
\end{equation}
For plaquette configurations $[0,0,0,0]$ or $[1,1,1,1]$ one always chooses 
time-like connections between sites, and for configurations $[0,1,1,0]$ or 
$[1,0,0,1]$ one always chooses space-like connections. For configurations 
$[0,1,0,1]$ or $[1,0,1,0]$ one chooses time-like connections with probability 
$p = A/(A+B) = 2/[1 + \exp(\epsilon/2)]$ and space-like connections with 
probability $1-p = B/(A+B)$. Indeed, this is the algorithm that was used in 
\cite{Wie94}. It is extremely efficient, has almost no detectable 
autocorrelations, and has a dynamical exponent for critical slowing down that
is compatible with zero. The cluster rules are illustrated in table 1.

\begin{table}[ht]
% space before first and after last column: 1.5pc
% space between columns: 3.0pc (twice the above)
%\setlength{\tabcolsep}{1.5pc}
% -----------------------------------------------------
% adapted from TeX book, p. 241
\newlength{\digitwidth} \settowidth{\digitwidth}{\rm 0}
\catcode`?=\active \def?{\kern\digitwidth}
% -----------------------------------------------------
  \begin{center}
    \leavevmode
%    \tiny
    \begin{tabular}{|c|c|}
	\hline
	configuration&break-ups\\
	\hline\hline
        \hspace{0.4cm}\begin{minipage}[c]{2.5cm}
		\epsfig{file=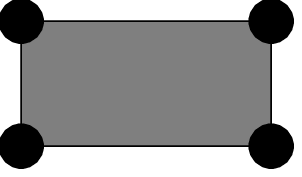,width=1.5cm,angle=270,
		bbllx=-5,bblly=0,bburx=55,bbury=85}
	\end{minipage}
	&\hspace{0.4cm}\begin{minipage}[c]{2.5cm}
		\vbox{\begin{center}
		\epsfig{file=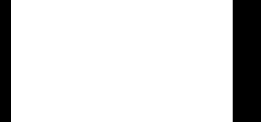,width=1.5cm,angle=270,
		bbllx=-20,bblly=0,bburx=55,bbury=85} 

		\hspace{-0.2cm}A
		\end{center}}
	\end{minipage} \\
	\hline
	\hspace{0.4cm}\begin{minipage}[c]{2.5cm}
		\epsfig{file=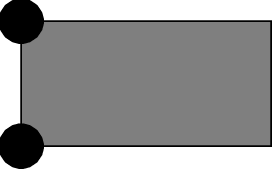,width=1.5cm,angle=270,
		bbllx=-5,bblly=0,bburx=55,bbury=85}
	\end{minipage}
	&$\begin{array}{c}
        \hspace{0.4cm}\begin{minipage}[c]{2.5cm}
		\vbox{\begin{center}
		\epsfig{file=fig/table_1_2.eps,width=1.5cm,angle=270,
		bbllx=-20,bblly=0,bburx=55,bbury=85} 

		\hspace{-0.2cm}\mbox{A}
		\end{center}}
	\end{minipage}
        \\
	\hspace{0.4cm}\begin{minipage}[c]{2.5cm}
		\vbox{\begin{center}
		\epsfig{file=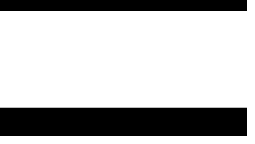,width=1.5cm,angle=270,
		bbllx=-20,bblly=0,bburx=55,bbury=85} 

		\hspace{-0.2cm}\mbox{B}
		\end{center}}
	\end{minipage}
        \end{array}$ \\
	\hline
	\hspace{0.4cm}\begin{minipage}[c]{2.5cm}
		\epsfig{file=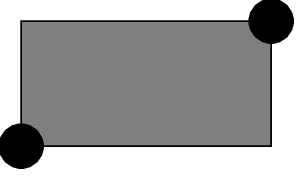,width=1.5cm,angle=270,
		bbllx=-5,bblly=0,bburx=55,bbury=85}
	\end{minipage}
	&\hspace{0.4cm}\begin{minipage}[c]{2.5cm}
		\vbox{\begin{center}
		\epsfig{file=fig/table_3_2.eps,width=1.5cm,angle=270,
		bbllx=-20,bblly=0,bburx=55,bbury=85} 

		\hspace{-0.2cm}B
		\end{center}}
	\end{minipage} \\
	\hline   
    \end{tabular}
\caption{\it Cluster break-ups of various plaquette configurations together 
with their relative probabilities $A$ and $B$. The dots represent occupied 
sites and the fat lines are the cluster connections.}
\end{center}
\end{table}

Eq.(\ref{cluster}) can be viewed as a representation of the original model as a
random cluster model. The cluster algorithm operates in two steps. First, a 
cluster break-up is chosen for each space-time interaction plaquette according 
to the above probabilities. This step effectively replaces the original 
Boltzmann 
weight of a plaquette configuration with a set of constraints represented by 
the $\delta$-functions associated with the chosen break-up. The constraints 
imply that occupation numbers of connected sites can only be changed together. 
Second, every cluster is flipped with probability 
$1/2$. When a cluster is flipped the occupation numbers of all sites that 
belong to the cluster are changed. Eq.(\ref{balance}) ensures that the cluster 
algorithm obeys detailed balance. 

Let us now consider the effect of a cluster flip on the fermion sign. Each 
cluster can be characterized by its effect on the fermion sign independent of 
the orientation of all the other clusters. We refer to clusters whose flip 
changes $\Sign[n]$ as merons, while clusters whose flip leaves $\Sign[n]$ 
unchanged are called non-merons. The flip of a meron-cluster permutes the 
fermions and changes the topology of the fermion world-lines. The number of 
merons in a configuration is always even. An example of a meron-cluster 
is given in figure 1. 

The meron concept alone allows us to gain an exponential factor in statistics.
Since all clusters can be flipped independently, one can construct an improved 
estimator for $\Sign[n]$ by averaging analytically over the $2^{N_C}$ 
configurations obtained by flipping the $N_C$ clusters in the configuration in 
all possible ways. For configurations that contain merons the average 
$\Sign[n]$ is zero because flipping a single meron leads to a cancellation of
contributions of $\pm 1$. Hence, only the configurations without merons 
contribute to $\langle \Sign \rangle$. The vast majority of configurations 
contains merons and now contributes an exact $0$ to $\langle \Sign \rangle$ 
instead of a statistical average of 
contributions $\pm 1$. In fact, one can show that the contributions from the
zero-meron sector are always positive, such that $\Sign[n]$ is $0$ for 
configurations containing meron-clusters and $1$ in the zero-meron sector.
According to the previous discussion, this method solves one half of the 
fermion sign 
problem. Before we can solve the other half of the problem we must discuss 
improved estimators for the physical observables.

Let us consider an improved estimator for $(\PP[n])^2 \Sign[n]$ which is needed
to determine the chiral susceptibility $\chi$. The total chiral condensate, 
$\PP[n] = \sum_C \PP_C$ is a sum of cluster contributions
\begin{equation} 
\PP_C = \frac{\epsilon}{6} \sum_{(x,t) \in C} (-1)^{x_1+x_2+x_3}
(n(x,t) - \frac{1}{2}).
\end{equation}
When a cluster is flipped, its condensate contribution changes sign. In a 
configuration without merons, where $\Sign[n] = 1$ for all relative cluster 
flips, the average of $(\PP[n])^2 \Sign[n]$ over all $2^{N_C}$ configurations 
is $\sum_C |\PP_C|^2$. For configurations with two merons the average is 
$2 |\PP_{C_1}||\PP_{C_2}|$ where $C_1$ and $C_2$ are the two meron-clusters. 
Configurations with more than two merons do not contribute to $(\PP[n])^2 
\Sign[n]$. The improved estimator for the susceptibility is hence given by
\begin{eqnarray}
\label{chi}
\chi = \frac{\langle \sum_C |\PP_C|^2 \delta_{N,0} + 2 |\PP_{C_1}||\PP_{C_2}| 
\delta_{N,2} \rangle}{V \beta \langle \delta_{N,0} \rangle}, \nonumber \\ \,
\end{eqnarray}
where $N$ is the number of meron-clusters in a configuration. Thus, to 
determine $\chi$ one must only sample the zero- and two-meron sectors. 

The probability to find a configuration without merons is exponentially small 
in the space-time volume since it is equal to $\langle \Sign \rangle$. Thus, 
although we have increased the statistics tremendously with the improved 
estimators, without a second step one would still need exponentially large 
statistics to accurately determine $\chi$. Fortunately, the numerator in
equation (\ref{chi}) receives contributions from the zero- and two-meron 
sectors only, while the denominator gets contributions only from the zero-meron
sector. One can hence restrict oneself to the zero- and two-meron sectors and 
never generate configurations with more than two merons. This enhances both the
numerator and the denominator by a factor that is exponentially large in the
volume, but leaves the ratio of the two invariant. One purpose of the second
step of the meron-cluster algorithm is to eliminate all configurations with 
more than two merons. To achieve this, we start with an initial configuration 
with zero or two merons. For example, a completely occupied configuration has 
no merons. We then visit all plaquette interactions one after the other and 
choose new pair connections between the four sites according to the above 
cluster rules. If the new connection increases the number of merons beyond two,
it is not accepted and the old connection is kept for that plaquette. This 
procedure obeys detailed balance because configurations with more than two 
merons do not contribute to the observable we consider. This simple reject step
eliminates almost all configurations with weight $0$ and is the essential step 
to solving the second half of the fermion sign problem. 

We have simulated the staggered fermion model with $G = 1$ on antiperiodic
spatial volumes $L^3$ with $L = 4,6,...,16$ and at various inverse temperatures
$\beta \in [0.5,1.2]$ which includes the critical point. In the Euclidean time
direction we have used $M = 4$, i.e. 24 time-slices. In all cases, we have 
performed at least 1000 thermalization sweeps followed by 10000 measurements.
One sweep consists of a new choice of the cluster connections on each 
interaction plaquette and a flip of each cluster with probability $1/2$.

\begin{figure}[htb]
\epsfig{file=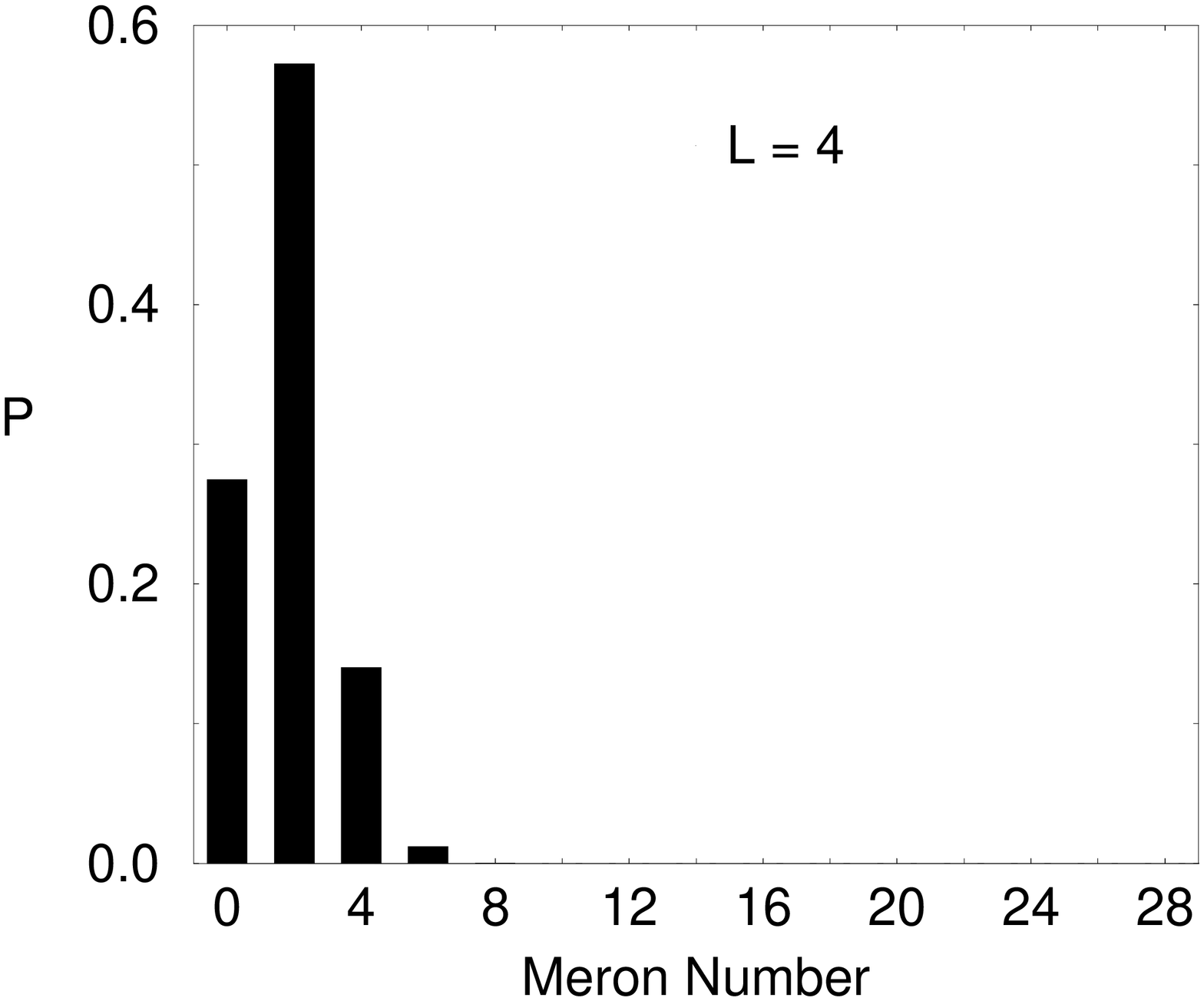,
width=6cm,angle=270,
bbllx=87,bblly=27,bburx=589,bbury=605}
\epsfig{file=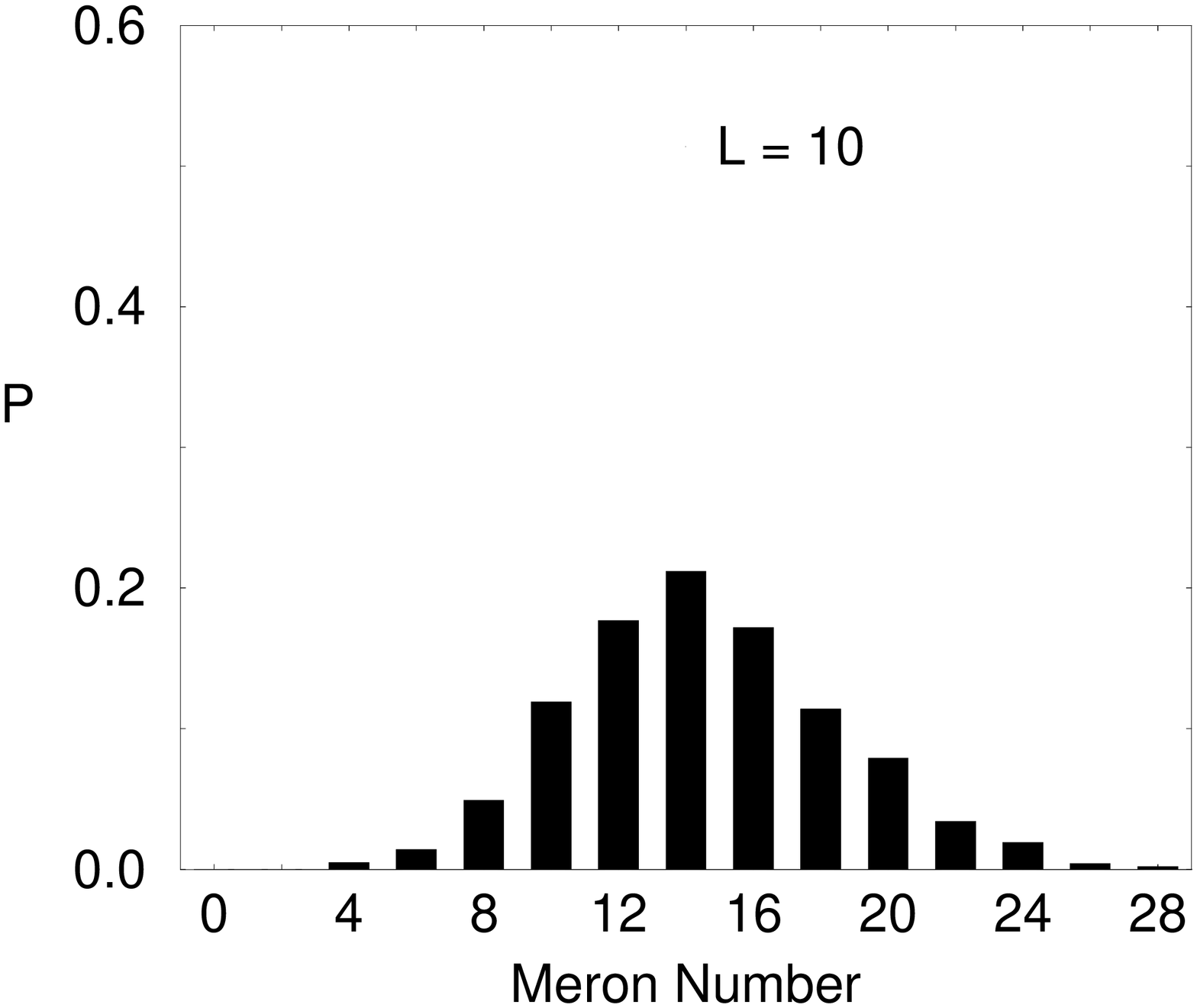,
width=6cm,angle=270,
bbllx=87,bblly=27,bburx=589,bbury=605}
\vspace{-1cm}
\caption{\it The probability of having a certain number of merons for
spatial size $L = 4$ and $10$ at $\beta = 0.948$.}
\end{figure}
Figure 3 shows the probability to have a certain number of merons in an 
algorithm that samples all meron-sectors. For small volumes the zero-meron 
sector and hence $\langle \Sign \rangle$ is relatively large, while multi-meron
configurations are rare. On the other hand, in larger volumes the vast majority
of configurations has a large number of merons and hence 
$\langle \Sign \rangle$ is exponentially small. For example, an extrapolation 
from smaller volumes gives a rough estimate 
$\langle \Sign \rangle \approx 10^{-20}$ on the $16^3$ lattice at 
$\beta = 0.948$. Hence, to achieve a similar accuracy without the meron-cluster
algorithm one would have to increase the statistics by a factor $10^{40}$, 
which is obviously impossible in practice. 

To study the critical behavior in detail, we have performed a finite-size
scaling analysis for $\chi$ focusing on a narrow range $\beta \in [0.9,0.98]$
around the critical point. Since a $\Z(2)$ chiral symmetry gets spontaneously 
broken at finite temperature in this $(3+1)$-d model, one expects to find the 
critical behavior of the 3-d Ising model. For the 3-d Ising model the critical 
exponents are given by $\nu = 0.630(1)$ and $\gamma/\nu = 1.963(3)$. Fitting to
our data, we find $\nu = 0.63(4)$ and $\gamma/\nu = 1.98(2)$, which indicates 
that the chiral transition of the staggered fermion model is indeed in the 3-d 
Ising universality class. In figure 3 we have taken the values of the critical 
exponents from the 3-d Ising model and we have plotted 
$\chi/L^{\gamma/\nu}$ as a function of $y = (\beta - \beta_c) L^{1/\nu}$. 
Indeed all data collapse onto one universal curve. 
\begin{figure}[htb]
\begin{center}
\epsfig{file=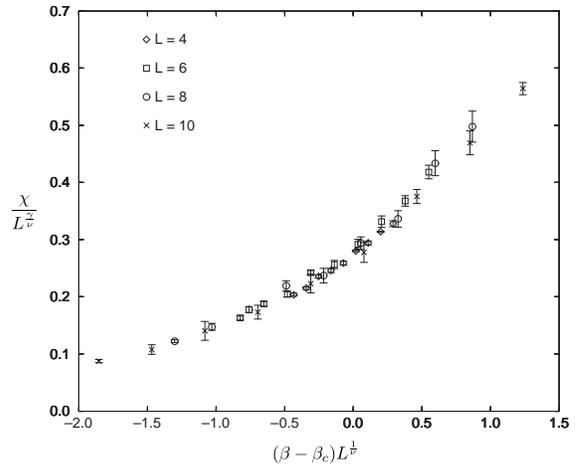,
width=7.5cm,angle=0,
bbllx=25,bblly=142,bburx=593,bbury=610}
\vspace{-1cm}
\caption{\it Finite-size scaling behavior of the chiral susceptibility $\chi$. 
The data for various spatial sizes $L = 4,6,8$ and $10$ fall on one universal 
curve.}
\end{center}
\end{figure}

\section{Quantum Antiferromagnets in a Magnetic Field$\,^4$}
\footnotetext[4]{Based on two talks given by B. Scarlet and U.-J. Wiese}

As another problem of physical interest we consider antiferromagnetic quantum 
spin ladders in an external uniform magnetic field that competes with the 
spin-spin interaction. The competing interactions lead to a severe sign 
problem in numerical simulations. The most efficient algorithm for simulating
quantum spin systems in the absence of an external magnetic field is the loop 
cluster algorithm \cite{Eve93,Wie94,Eve97} which can be implemented directly in
the Euclidean time continuum \cite{Bea96}. This algorithm also underlies our
staggered fermion simulations and was described in detail in section 3. In the 
presence of a magnetic field pointing along the quantization axis of the spins,
the $\Z(2)$ symmetry that allows the clusters to be flipped with probability 
$1/2$ is explicitly broken. As a consequence, for large values of the magnetic 
field some clusters can only be flipped with very small probability and the 
algorithm becomes inefficient. An alternative strategy is to choose the spin 
quantization axis perpendicular to the direction of the magnetic field. Then 
all clusters can still be flipped with probability $1/2$. This indeed leads to 
an efficient algorithm for ferromagnets in an external uniform magnetic field. 
Unfortunately, for antiferromagnets i.e. when the magnetic field competes with
the spin-spin interaction --- this formulation of the problem leads to a very 
severe sign problem. Here we show how this sign problem can be solved 
completely using a meron-cluster algorithm.

Antiferromagnetic spin ladders --- sets of several transversely coupled quantum
spin chains --- are interesting condensed matter systems which interpolate 
between single 1-d spin chains and 2-d quantum antiferromagnets. The ladders 
are spatially quasi 1-d systems whose low-energy dynamics are governed by 
$(1+1)$-d quantum field theories. We consider spin ladders in an external 
uniform magnetic field $B$, which corresponds to a chemical potential 
$\mu = B/c$ in the corresponding $(1+1)$-d quantum field theory. Chakravarty, 
Halperin and Nelson used a $(2+1)$-d effective field theory to describe the 
low-energy dynamics of spatially 2-d quantum antiferromagnets \cite{Cha88}. 
Chakravarty has applied this theory to quantum spin ladders with a sufficiently
large even number of coupled spin $1/2$ chains \cite{Cha96}. We consider spin 
ladders with the same value of the antiferromagnetic coupling along and between
the chains. These systems are described by the action
\begin{eqnarray}
S[\vec e]&=&\int_0^\beta dt \int_0^L dx \int_0^{L'} dy \
\frac{\rho_s}{2}[\partial_x \vec e \cdot \partial_x \vec e \nonumber \\
&+&\partial_y \vec e \cdot \partial_y \vec e + 
\frac{1}{c^2} \partial_t \vec e \cdot \partial_t \vec e].
\end{eqnarray}
Here $\vec e(x,y,t)$ is a unit-vector field, $\rho_s$ is the spin stiffness and
$c$ is the spin-wave velocity. The coupled spin chains are oriented in the 
spatial $x$-direction with a large extent $L$, while the transverse 
$y$-direction has a much smaller extent $L' \ll L$. Here we consider spin 
ladders with periodic boundary conditions in the transverse direction, and we 
limit ourselves to an even number of coupled spin $1/2$ chains. The effective 
action for a ladder with an odd number of coupled chains would contain an 
additional topological term.

When the spin ladder is placed in a uniform external magnetic field $\vec B$,
the field couples to a conserved quantity --- the uniform magnetization. Hence,
on the level of the effective theory, the magnetic field plays the role of a
chemical potential --- i.e. it appears as the time-component of an imaginary
constant vector potential. The ordinary derivative
$\partial_t \vec e$ is replaced by the covariant derivative $\partial_t \vec e
+ i \vec B \times \vec e$. For a sufficiently large even number of coupled 
chains ($L' \gg c/\rho_s$) the ladder system undergoes dimensional reduction to
the $(1+1)$-d $O(3)$ symmetric quantum field theory with the action
\begin{eqnarray}
S[\vec e]&=&\int_0^\beta dt \int_0^L dx \
\frac{\rho_s L'}{2}[\partial_x \vec e \cdot \partial_x \vec e \nonumber \\
&+&\frac{1}{c^2} (\partial_t \vec e + i \vec B \times \vec e) \cdot
(\partial_t \vec e + i \vec B \times \vec e)].
\end{eqnarray}
The effective coupling constant is given by $1/g^2 = \rho_s L'/c$ and the 
magnetic field appears as a chemical potential of magnitude $\mu = B/c$.

We consider a system of quantum spins $1/2$ on a $d$-dimensional cubic lattice 
with site label $x$ and with periodic spatial boundary conditions. In 
particular, we are interested in ladder systems on a 2-d rectangular lattice of
size $L \times L'$ with $L \gg L'$. The spins located at the sites $x$ are 
described by operators $S^i_x$ with the usual commutation relations
\begin{equation}
[S_x^i,S_y^j] = i \delta_{xy} \epsilon_{ijk} S_x^k.
\end{equation}
The Hamilton operator 
\begin{equation}
H = J \sum_{x,i} \vec S_x \cdot \vec S_{x+\hat i} - 
\vec B \cdot \sum_x \vec S_x,
\end{equation}
with $J > 0$, couples the spins at the lattice sites $x$ and $x+\hat i$, where 
$\hat i$ is a unit-vector in the $i$-direction. A path integral representation 
of the partition function can be derived in complete analogy to the staggered
fermion model of section 3. In this case, one sums over configurations of spins
$s(x,t) = \up, \down$ on a $(d+1)$-dimensional space-time lattice of points 
$(x,t)$. Again, the Boltzmann factor is a product of space-time plaquette 
contributions with $\Sign[s(x,t),s(x+\hat i,t),s(x,t+1),s(x+\hat i,t+1)]$ given
by
\begin{eqnarray}
&&\Sign[\up,\up,\up,\up] = \Sign[\up,\down,\up,\down] = 
\nonumber \\
&&\Sign[\down,\up,\down,\up] = \Sign[\down,\down,\down,\down] = 1
\nonumber \\
&&\Sign[\up,\down,\down,\up] = \Sign[\down,\up,\up,\down] = - 1.
\end{eqnarray}
When the magnetic field points perpendicular to the spin quantization axis,
time-like bond contributions to the Boltzmann factor also arise which have
\begin{eqnarray}
\label{Boltzmann2}
&&e^{- S[\up,\up]} = e^{- S[\down,\down]} = \cosh(\epsilon B/2) \nonumber \\
&&e^{- S[\up,\down]} = e^{- S[\down,\up]} = \sinh(\epsilon B/2).
\end{eqnarray}
Figure 4 shows two spin configurations in $(1+1)$ dimensions. The first
configuration is completely antiferromagnetically ordered and has 
$\Sign[s] = 1$. The second configuration contains one interaction plaquette
with configuration $[\down,\up,\up,\down]$ which contributes 
$\Sign[\down,\up,\up,\down] = - 1$. In addition, there are two time-like 
interaction bonds with configurations $[\down,\up]$ and $[\up,\down]$. 
\begin{figure}[htb]
\epsfig{file=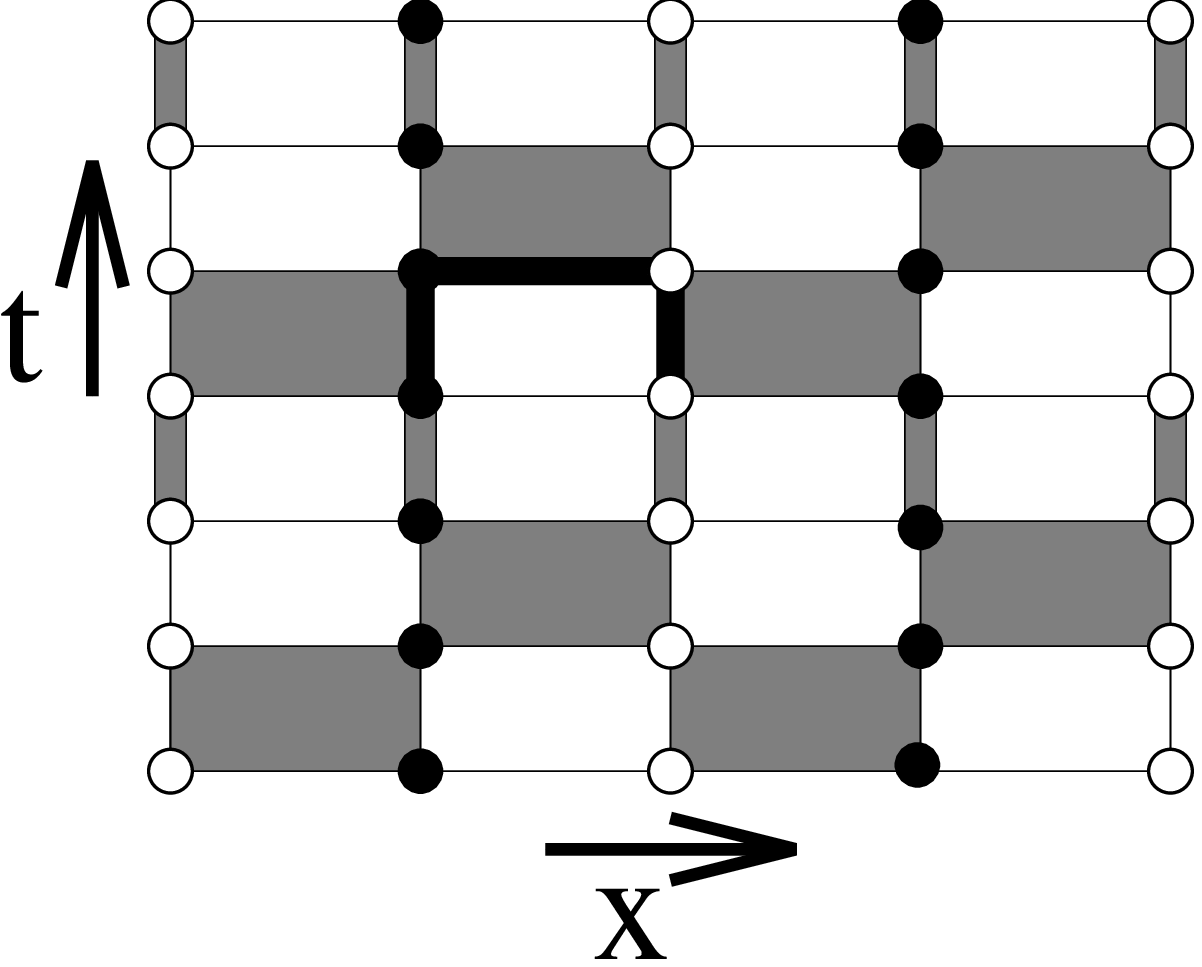,
height=5.5cm,angle=270,
bbllx=0,bblly=0,bburx=276,bbury=294}
\epsfig{file=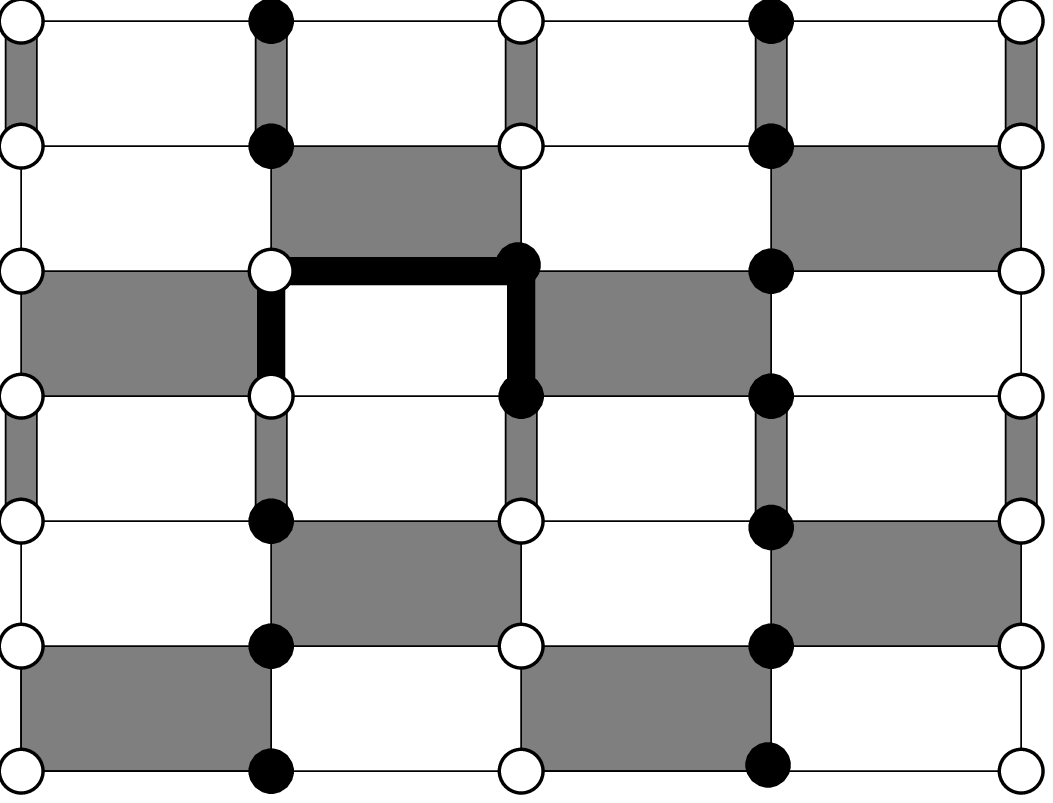,
height=5.5cm,angle=270,
bbllx=0,bblly=-44,bburx=230,bbury=250}
\vspace{-0.5cm}
\caption{\it Two spin configurations in $(1+1)$ dimensions. The shaded
plaquettes and shaded time-like bonds carry the interaction. Filled dots 
represent 
spin up and open circles represent spin down. The second configuration has 
$\Sign[s] = - 1$. A meron-cluster is represented by the fat black line.
Flipping this cluster changes one configuration into the other and changes 
$\Sign[s]$.}
\end{figure}

The central observable of our study is the uniform magnetization
\begin{equation}
\vec M = \sum_x \vec S_x.
\end{equation}
The expectation value of the magnetization $\langle M \rangle$ in the 
direction of the magnetic field is non-zero.

The cluster algorithm for the quantum antiferromagnet is very similar to the
one for staggered fermions. Just like the space-time plaquette terms, the 
time-like bond Boltzmann factors are expressed as
\begin{equation}
\label{cluster2}
e^{- S[s(x,t),s(x,t+1)]} = C \delta_{s(x,t),s(x,t+1)} + D.
\end{equation}
The probability to connect spins with their time-like neighbors is $C/(C+D)$.
The spins remain disconnected with probability $D/(C+D)$. Inserting the 
expressions from eq.(\ref{Boltzmann2}) one obtains
\begin{equation}
\label{balance2}
C + D = \cosh(\epsilon B/2), C = \sinh(\epsilon B/2).
\end{equation}
The cluster rules for the space-time plaquette terms are the same as in table 
1. The rules for the time-like bonds are illustrated in table 2.
\begin{table}[htb]
% space before first and after last column: 1.5pc
% space between columns: 3.0pc (twice the above)
%\setlength{\tabcolsep}{1.5pc}
% -----------------------------------------------------
% adapted from TeX book, p. 241
%\newlength{\digitwidth} \settowidth{\digitwidth}{\rm 0}
\catcode`?=\active \def?{\kern\digitwidth}
% -----------------------------------------------------
  \begin{center}
    \leavevmode
%    \tiny
    \begin{tabular}{|c|c|}
	\hline
	configuration&break-ups\\
	\hline\hline
	\hspace{1.4cm}\begin{minipage}[c]{1.5cm}
		\epsfig{file=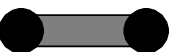,width=1.5cm,angle=270,
		bbllx=-5,bblly=0,bburx=55,bbury=85}
	\end{minipage}
	&\hspace{0.4cm}\begin{minipage}[c]{0.5cm}
		\vbox{\begin{center}
		\epsfig{file=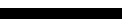,width=1.5cm,angle=270,
		bbllx=-10,bblly=-5,bburx=45,bbury=5} 

		\hspace{-0.0cm}C
		\end{center}}
	\end{minipage}
	\hspace{2.4cm}\begin{minipage}[c]{0.5cm}
		\vbox{\begin{center}
		\epsfig{file=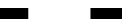,width=1.5cm,angle=270,
		bbllx=-10,bblly=-5,bburx=45,bbury=5} 

		\hspace{-0.0cm}D
		\end{center}}
	\end{minipage}
	\\
	\hline   
	\hspace{1.4cm}\begin{minipage}[c]{1.5cm}
		\epsfig{file=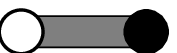,width=1.5cm,angle=270,
		bbllx=-5,bblly=0,bburx=55,bbury=85}
	\end{minipage}
	&\hspace{0.4cm}\begin{minipage}[c]{0.5cm}
		\vbox{\begin{center}
		\epsfig{file=fig/table_4_3.eps,width=1.5cm,angle=270,
		bbllx=-10,bblly=-5,bburx=45,bbury=5} 

		\hspace{-0.0cm}D
		\end{center}}
	\end{minipage}
	\\
	\hline   
    \end{tabular}
\caption{\it Cluster break-ups of time-like bond configurations together with 
their probabilities $C$ and $D$. Filled dots represent spin up, open circles 
represent spin down, and the fat black line is the cluster connection.}
\end{center}
\end{table}

The above cluster rules were first used in a simulation of the Heisenberg 
antiferromagnet \cite{Wie94} in the absence of a magnetic field. In that case
there is no sign problem. Then the corresponding loop-cluster algorithm is 
extremely efficient and has almost no detectable autocorrelations. When a 
magnetic field is switched on the situation changes. When the magnetic field 
points in the direction of the spin quantization axis (the $3$-direction in our
case) there is no sign problem. However, the magnetic field then explicitly 
breaks the $\Z(2)$ flip symmetry on which the cluster algorithm is based, and 
the clusters can no longer be flipped with probability $1/2$. Instead the flip 
probability is determined by the value of the magnetic field and by the 
magnetization of the cluster. When the field is strong, flips of magnetized
clusters are rarely possible and the algorithm becomes very inefficient. To 
avoid this, we have chosen the magnetic field to point perpendicular to the 
spin quantization axis. In that case, the cluster flip symmetry is not affected
by the magnetic field, and the clusters can still be flipped with probability 
$1/2$. However, now a severe sign problem arises. 

One can construct an improved estimator for the magnetization 
$\langle M \rangle$, which gets non-vanishing contributions only from the 
zero-meron sector. Hence, it is unnecessary to generate any configurations that
contain meron-clusters. In practice it is advantageous to occasionally 
generate configurations containing merons even though they do not contribute to
our observable, because this reduces the autocorrelation times. 

\begin{figure}[htb]
\epsfig{file=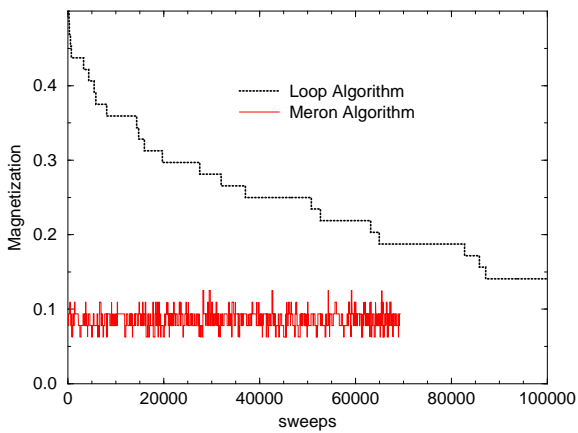,
width=8.5cm,angle=0,
bbllx=42,bblly=20,bburx=602,bbury=445}
\vspace{-1.5cm}
\caption{\it Thermalization of the magnetization for the loop-cluster algorithm
versus the meron-cluster algorithm. For $B = J$ the loop algorithm takes more 
than 100000 sweeps to reach equilibrium while the meron-cluster algorithm has 
no thermalization problem.}
\end{figure}
We have performed numerical simulations with the meron-cluster algorithm for
various quantum antiferromagnets in a uniform magnetic field $B$. To 
demonstrate the efficiency of the algorithm, we have compared it with the
standard loop-cluster algorithm. In case of the loop algorithm the magnetic
field points in the direction of the spin quantization axis. In the 
meron-cluster algorithm, on the other hand, the magnetic field is perpendicular
to the spin quantization axis and all clusters can still be flipped with 
probability $1/2$. Of course the sign problem arises, but it is solved 
completely by the meron-cluster algorithm. Figure 5 compares the thermalization
behavior of the magnetization of a 2-d Heisenberg antiferromagnet on an 
$8 \times 8$ lattice at $\beta J = 10$ with $M = 100$ for the loop-cluster 
algorithm and the meron-cluster algorithm. At $B=J$ the loop-cluster algorithm 
needs more than 100000 equilibration sweeps, while the meron-cluster algorithm 
has no thermalization problem. 

\begin{figure}[htb]
\epsfig{file=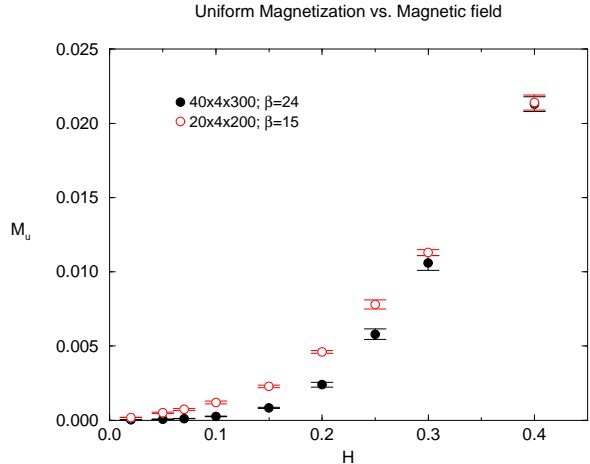,
width=8.5cm,angle=0,
bbllx=43,bblly=21,bburx=605,bbury=447}
\vspace{-1.5cm}
\caption{\it Magnetization density as a function of the magnetic field.}
\end{figure}
Figure 6 shows the magnetization density of antiferromagnetic quantum spin 
ladders consisting of four coupled spin $1/2$ chains (i.e. $L'/a = 4$). The
cases $L = 20, \beta J = 15, M = 200$ and $L = 40, \beta J = 24, M = 300$ have
been considered. In the infinite volume, zero temperature limit one expects the
magnetization to vanish for $B < B_c$. The critical magnetic field $B_c$ 
corresponds to the critical chemical potential $\mu = B_c/c = m$, determined by
the mass gap $m$ of the $(1+1)$-d $O(3)$ model. The expected behavior is indeed
consistent with the numerical data.
 
\section{Strongly correlated Electron Systems$\,^3$}
\footnotetext[3]{Based on a talk presented by J. Cox}

Quantum Antiferromagnets may turn into high-temperature superconductors when
they are doped with additional charge carriers. These systems are believed to
be well described by the Hubbard model Hamiltonian
\begin{eqnarray}
H&=&- \frac{t}{2} \sum_{x,i,s=\up,\down}(\Psi^+_{x,s} \Psi_{x+\hat i,s} +
\Psi^+_{x+\hat i,s} \Psi_{x,s}) \nonumber \\
&+&U \sum_x (n_{x,\up} - \frac{1}{2})(n_{x,\down} - \frac{1}{2}) - 
\mu \sum_x (n_x - 1). \nonumber \\ \,
\end{eqnarray}
Here $\Psi^+_{x,s}$ and $\Psi_{x,s}$ are creation and annihilation operators
for electrons with spin $s=\up,\down$ hopping on the sites $x$ of a quadratic 
lattice. The chemical potential $\mu$ couples to the occupation number per site
$n_x = n_{x,\up} + n_{x,\down} = \Psi^+_{x,\up} \Psi_{x,\up} + 
\Psi^+_{x,\down} \Psi_{x,\down}$ and is used to dope the system away from 
half-filling. So far, high-temperature superconductivity has not been 
demonstrated convincingly in numerical simulations due to a very severe fermion
sign problem. It is natural to ask if a meron-cluster algorithm can be used to 
solve this problem. Unfortunately, as it stands the meron concept does not 
apply directly to the Hubbard model because the clusters influence each other 
in their effect on the fermion sign. 

Meron-cluster algorithms can solve the sign problem when two conditions are
satisfied. First, the clusters must be independent in their effect on the sign
in order to allow the construction of improved estimators. Second, it must 
always be possible to flip the clusters into a reference configuration with a 
positive sign. In case of the Hubbard model a completely antiferromagnetically 
ordered reference configuration suggests itself. In order to construct an 
efficient meron-cluster algorithm based on that reference configuration, one 
must modify the original Hubbard model Hamiltonian. Of course, one would like 
to maintain the symmetry properties of the original Hamiltonian in order to 
stay in the same universality class. An obvious symmetry is the $SU(2)_s$ spin
rotational symmetry that is generated by 
$\vec S = \frac{1}{2} \sum_{x,s} \Psi^+_{x,s} \vec \sigma_{ss'} \Psi_{x,s'}$.
Another less obvious $SU(2)_n$ symmetry is generated by $\vec N$ with
\begin{eqnarray}
N^+&=&\sum_x (-1)^{x_1+x_2} \Psi^+_{x,\up} \Psi^+_{x,\down}, N^- = (N^+)^+,
\nonumber \\
N^3&=&\frac{1}{2} \sum_x (\Psi^+_{x,\up} \Psi_{x,\up} + 
\Psi^+_{x,\down} \Psi_{x,\down}).
\end{eqnarray}
We have systematically investigated the space of $SU(2)_s \otimes SU(2)_n$
nearest neighbor interaction Hamiltonians for which an efficient meron-cluster
algorithm can be constructed. One example of such a Hamiltonian is
\begin{eqnarray}
H&=&- \frac{t}{2} \sum_{x,i,s=\up,\down}(\Psi^+_{x,s} \Psi_{x+\hat i,s} +
\Psi^+_{x+\hat i,s} \Psi_{x,s}) \nonumber \\
&+&\frac{t'}{2} \sum_{x,i,s=\up,\down}(\Psi^+_{x,s} \Psi_{x+\hat i,s} +
\Psi^+_{x+\hat i,s} \Psi_{x,s}) \nonumber \\
&\times&(n_x + n_{x+\hat i} - 2)^2 \nonumber \\
&+&J \sum_{x,i} \vec S_x \cdot \vec S_{x+\hat i} \nonumber \\
&+&U \sum_x (n_{x,\up} - \frac{1}{2})(n_{x,\down} - \frac{1}{2}) \nonumber \\
&+&V \sum{x,i} [(n_x - 1)^2 + (n_{x+\hat i} - 1)^2 \nonumber \\
&-&2(n_x - 1)^2 (n_{x+\hat i} - 1)^2] - \mu \sum_x (n_x - 1).
\end{eqnarray} 
In particular, an additional antiferromagnetic coupling 
$J \vec S_x \cdot \vec S_{x+\hat i}$ arises, which is constrained by 
$J \geq 2t$. Furthermore, $t' = t$, $U \geq J$ and $U \geq 8V$. Indeed the 
above system can be simulated reliably with a very efficient meron-cluster 
algorithm. Figure 7 shows the density of charge carriers away from half-filling
as a function of the chemical potential for various temperatures. At $\mu 
\approx 0.45$ there is a strong first order phase transition at which the 
system undergoes phase separation. So far we have not found 
high-temperature superconductivity. This result may be a consequence of the 
additional
antiferromagnetic coupling. However, it could also indicate that even the
original Hubbard model does not display high-temperature superconductivity.
\begin{figure}[htb]
\epsfig{file=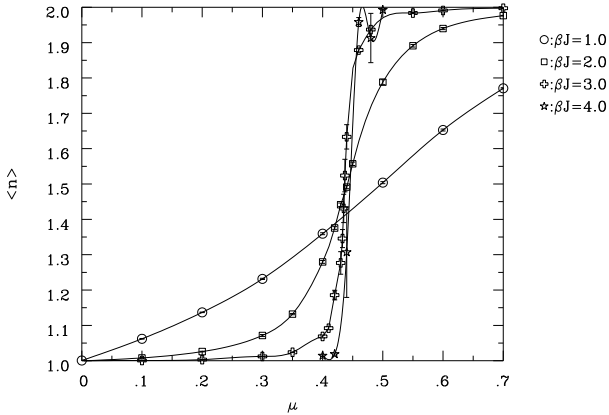,
width=5.5cm,angle=90,
bbllx=70,bblly=97,bburx=534,bbury=761}
\vspace{-1.5cm}
\caption{\it Particle density as a function of chemical potential for four
different temperatures.}
\end{figure}

\section{The 2-d $O(3)$ Model at non-zero Chemical Potential$\,^4$}
\footnotetext[4]{Based on two talks given by B. Scarlet and U.-J. Wiese}

Another severe sign problem arises in simulations of field theories at non-zero
chemical potential. Obviously, understanding QCD at non-zero baryon density is
of great experimental and theoretical importance. Here we study a simpler toy
model for QCD --- the 2-d $O(3)$ model. Like QCD, this model is asymptotically
free and has a non-perturbatively generated mass gap, instantons and 
$\theta$ vacua. In addition, it also suffers from a very severe sign problem 
when coupled to a chemical potential. As already discussed in section 4, the 
action of the 2-d $O(3)$ model with chemical potential $\vec \mu$ takes the 
form
\begin{eqnarray}
\label{O3action}
S[\vec e]&=&\int d^2x \frac{1}{2 g^2} 
[\partial_1 \vec e \cdot \partial_1 \vec e \nonumber \\
&+&(\partial_2 \vec e + i \vec \mu \times \vec e) \cdot 
(\partial_2 \vec e + i \vec \mu \times \vec e)].
\end{eqnarray}
Note that in this case the chemical potential is a vector because it couples to
a non-Abelian conserved charge.

Due to the use of discrete variables, a meron-cluster algorithm can be applied 
to the D-theory formulation of the 2-d $O(3)$ model. Instead of performing a 
path integral over classical fields, in the D-theory formulation of field 
theory \cite{Cha97,Bro97,Bea98,Wie99} collective excitations of discrete 
quantum variables which undergo dimensional reduction play the role of an 
effective classical field. For example, in D-theory the 2-d $O(3)$ model arises
from dimensional reduction of the $(2+1)$-d antiferromagnetic quantum 
Heisenberg model with the Hamiltonian
\begin{equation}
H = J \sum_{x,i} \vec S_x \cdot \vec S_{x+\hat i}.
\end{equation}
At zero temperature this model develops a staggered magnetization and the 
$SO(3)$ spin rotational symmetry breaks spontaneously down to $SO(2)$. The
low-energy dynamics of the corresponding Goldstone bosons --- the magnons or
spin waves --- are described by the chiral perturbation theory action
\begin{equation}
S[\vec e] = \int_0^\beta dt \int d^2x \frac{\rho_s}{2} 
[\partial_i \vec e \cdot \partial_i \vec e + 
\frac{1}{c^2} \partial_t \vec e \cdot \partial_t \vec e].
\end{equation}
Here $\rho_s$ is the spin stiffness and $c$ is the spin wave velocity. At low
temperatures the correlation length of the system diverges as $\xi \propto
\exp(2 \pi \rho_s \beta)$. Consequently, the extent $\beta$ of the Euclidean
time direction becomes negligible compared to $\xi$ and the system undergoes
dimensional reduction. The dimensionally reduced theory is the 2-d $O(3)$ model
\begin{equation}
S[\vec e] = \int d^2x \frac{\rho_s \beta}{2} 
\partial_i \vec e \cdot \partial_i \vec e
\end{equation}
without chemical potential. In order to incorporate a non-zero 
chemical potential, the D-theory Hamiltonian must be modified to
\begin{eqnarray}
\label{Hchemical}
H&=&J \sum_x [\vec S_x \cdot \vec S_{x+\hat 1} \nonumber \\
&+&\frac{1}{2}(S^+_x S^-_{x+\hat 2} e^\mu +
S^-_x S^+_{x+\hat 2} e^{-\mu}) + S^3_x S^3_{x+\hat 2}].
\end{eqnarray}
This system dimensionally reduces to the 2-d $O(3)$ model with chemical 
potential of eq.(\ref{O3action}).

The above Hamiltonian can be simulated with a meron-cluster algorithm very much
like the one for quantum antiferromagnets in a magnetic field. The resulting
particle density as a function of the chemical potential is shown in figure 8.
\begin{figure}[htb]
\vspace{-0.7cm}
\epsfig{file=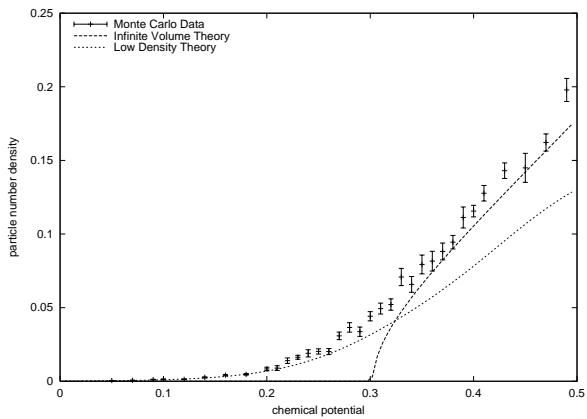,
width=5.5cm,angle=270,
bbllx=52,bblly=50,bburx=552,bbury=769}
\vspace{-0.8cm}
\caption{\it The particle density as a function of the chemical potential in
the 2-d $O(3)$ model. The Monte Carlo data are consistent with analytic
predictions at low density and at infinite volume represented by the dashed
lines.}
\end{figure}
The Monte Carlo results show the expected onset behavior when the chemical 
potential exceeds the mass gap $m = 1/\xi$. In the zero temperature limit no 
particles would be produced for $\mu < m$. At finite temperatures, on the other
hand, a small particle density which can be computed analytically exists in 
this region. As shown in figure 8, the Monte Carlo data are consistent with 
this analytic prediction. For larger values of $\mu$ finite temperature effects
are small and one can compare with the thermodynamic Bethe ansatz solution of
the 2-d $O(3)$ model, which again is consistent with the numerical data.

The steps taken here for the 2-d $O(3)$ model can potentially all be 
generalized to QCD. In the D-theory formulation, QCD is described as a quantum
link model \cite{Bro97}, and a chemical potential can be included exactly as in
eq.(\ref{Hchemical}). The quarks in quantum link QCD appear as domain wall
fermions. Hence, to simulate QCD with chemical potential along these lines, one
must first construct a meron-cluster algorithm for domain wall fermions.

\section{Domain Wall Fermions$\,^5$}
\footnotetext[5]{Based on a talk presented by C. Gattringer}

For simplicity, let us consider the Hamiltonian of free $(2+1)$-d domain wall
fermions, which describes 2-d chiral fermions bound to the walls
\begin{eqnarray}
\label{QCDaction}
H&=&\frac{1}{2} \sum_{x,i} [\Psi^\dagger_x \gamma_3 \gamma_i \Psi_{x+\hat i} 
- \Psi^\dagger_{x+\hat i} \gamma_3 \gamma_i \Psi_x] 
\nonumber \\
&+&\frac{r}{2} \sum_{x,i} \ [2 \Psi^\dagger_x \gamma_3 \Psi_x 
- \Psi^\dagger_x \gamma_3 \Psi_{x+\hat i}
- \Psi^\dagger_{x+\hat i} \gamma_3 \Psi_x] \nonumber \\
&+&M \sum_x \Psi^\dagger_x \gamma_3 \Psi_x.
\end{eqnarray}
Here $\Psi^\dagger_x$ and $\Psi_x$ are quark creation and annihilation 
operators obeying canonical anticommutation relations. Following \cite{Fur95}, 
the partition function is written as
\begin{equation}
Z = \langle 0|\exp(- \beta H)|0\rangle,
\end{equation}
where $|0\rangle$ is a particular fermion Fock state. Taking the expectation
value in that state implies that there are no left-handed quarks at $x_3 = 0$, 
and no right-handed quarks at $x_3 = \beta$.

Like for the Hubbard model, we have thus far not found an efficient 
meron-cluster algorithm for the original domain wall fermion Hamiltonian.
Instead, we ask for which Hamiltonians an efficient algorithm can be 
constructed. In order to stay in the same universality class, we demand that 
these Hamiltonians have the same symmetry properties as the original domain
wall fermion Hamiltonian. The relevant symmetries are charge conjugation C,
parity P, and lattice rotations R by 90 degrees. As in the staggered fermion 
case of section 3, we decompose the Hamiltonian as $H = \sum_{x,i} h_{x,i}$ and
we consider the transfer matrix $T_i = \exp(- \epsilon h_{x,i})$. In the
Hilbert space, the symmetries C, P and R are represented by unitary 
transformations $U_C$, $U_P$ and $U_R$. The symmetry requirements on the
$(2+1)$-dimensional transfer matrix thus take the form
\begin{equation}
T_i = U^+_{C,P} T_i U_{C,P}, \ T_2 = U^+_R T_1 U_R.
\end{equation}
In order to stay in the right universality class, it is equally important not
to have any additional symmetries that are broken in the original domain wall 
fermion Hamiltonian. This is particularly important for removing doubler 
fermions. Like for the Hubbard model, we have systematically investigated all
symmetry requirements, and we have identified the most general nearest neighbor
interaction Hamiltonian for which an efficient meron-cluster algorithm can be 
constructed. Compared to previous cases an additional complication arises. The
$\gamma$-matrices give rise to contributions $\pm i$ to the total Boltzmann 
weight. Hence, the flip of a general cluster may not only change the sign ---
it may change the complex phase of the Boltzmann weight by a factor $i$. 
Consequently, besides the sign-changing meron-clusters there are i-on clusters 
whose flip changes the phase of the Boltzmann weight by a factor $i$.

At present, we have not fully explored the space of efficient cluster 
algorithms for domain wall fermions. The necessary modifications of the 
standard Hamiltonian are similar to the ones required in the Hubbard model. It
remains to be seen if chiral fermions bound to the walls appear in the modified
models that can be simulated with efficient meron-cluster algorithms.

\section{Conclusion}

In conclusion, the meron concept provides us with a powerful algorithmic tool
--- the meron-cluster algorithm --- which can lead to a complete solution of
severe sign problems. Here we have demonstrated that this algorithm allows us 
to simulate staggered fermions with an unusually small number of flavors,
certain strongly correlated electron systems, quantum antiferromagnets in a
magnetic field, as well as the $O(3)$ model at non-zero chemical potential.
The next challenge is to construct meron-cluster algorithms for QCD at non-zero
baryon density and for systems that show high-temperature superconductivity. 
 
\section*{Acknowledgments}

We like to thank R. Brower and S. Chandrasekharan, who is involved in all 
projects discussed here, for a very pleasant and productive collaboration on 
meron-cluster algorithms. We also thank P. Hasenfratz and F. Niedermayer for
very interesting discussions. The work described here is supported in part by 
funds provided by the U.S. Department of Energy (D.O.E.) under cooperative 
research agreement DE-FC02-94ER40818. U.-J. W. also would like to thank the 
A. P. Sloan foundation for its support.

\end{document}